\documentclass[aps,prl,twocolumn,amsmath,amssymb,showpacs]{revtex4-1}
\usepackage{graphicx}
\usepackage{dcolumn}
\usepackage{bm}

\usepackage{latexsym}
\usepackage{amsthm}
\usepackage{braket}
\allowdisplaybreaks

\usepackage{enumitem}

\usepackage[top = 1. in,bottom = 1. in, left = 1 in, right=1 in]{geometry}

 \newcommand{\arXiv}[1]{\href{http://www.arXiv.org/abs/#1}{arXiv:#1}}
\usepackage[colorlinks=true, linkcolor=blue, bookmarks=true]{hyperref}

\makeatletter
\renewcommand\section{\@startsection {section}{1}{\z@}%
                  {-3.5ex \@plus -1ex \@minus -.2ex}
                  {2.3ex \@plus.2ex}%
                  {\normalfont\large\bfseries}}
\renewcommand\subsection{\@startsection{subsection}{2}{\z@}%
                   {-3.25ex\@plus -1ex \@minus -.2ex}%
                   {1.5ex \@plus .2ex}%
                   {\normalfont\bfseries}}
\makeatother

\newcommand{\mm}{\mu}
\newcommand{\tildeC}{S}

\newcommand{\beq}{\begin{equation}}
\newcommand{\eeq}{\end{equation}}
\newcommand{\ber}{\begin{array}}
\newcommand{\eer}{\end{array}}
\newcommand{\del}{\partial}
\newcommand{\sssty}{\scriptscriptstyle}
\newcommand{\ssty}{\scriptstyle}

\newcommand{\ena}{\end{eqnarray}}
\newcommand{\beqa}{\begin{eqnarray}}
\newcommand{\eeqa}{\end{eqnarray}}
\newcommand{\bea}{\begin{eqnarray}}
\newcommand{\eea}{\end{eqnarray}}

\theoremstyle{remark}

\renewcommand{\Re}{\operatorname{Re}}
\renewcommand{\Im}{\operatorname{Im}}

\begin{document}

\title{Two infinite families of resonant solutions for the Gross-Pitaevskii equation}
\author {Anxo Biasi,$^{1}$ Piotr Bizo\'n,$^{2}$ Ben Craps,$^{3}$ Oleg Evnin$^{4,3}$\vspace{2mm}}

\affiliation{ $^{1}$Departamento de F\'\i sica de Part\'\i culas, Universidade de Santiago de Compostela
	 and Instituto Galego de F\'\i sica de Altas Enerx\'\i as (IGFAE), Santiago de Compostela 15782, Spain\\
 $^{2}$Institute of Physics, Jagiellonian University, Krak\'ow 30-348, Poland\\
$^{3}$Theoretische Natuurkunde, Vrije Universiteit Brussel (VUB) and
	The International Solvay Institutes, B-1050 Brussels, Belgium\\
$^{4}$Department~of~Physics,~Faculty~of~Science,~Chulalongkorn~University,~Bangkok 10330,~Thailand}

\begin{abstract}
We consider the two-dimensional Gross-Pitaevskii equation describing a Bose-Einstein condensate in an isotropic harmonic trap. In the small coupling regime, this equation is accurately approximated over long times by the corresponding nonlinear resonant system whose structure is determined by the fully resonant spectrum of the linearized problem. We focus on two types of consistent truncations of this resonant system: first, to sets of modes of fixed angular momentum, and second, to excited Landau levels. Each of these truncations admits a set of explicit analytic solutions with initial conditions parametrized by three complex numbers. Viewed in position space, the fixed angular momentum solutions describe modulated oscillations of dark rings, while the excited Landau level solutions describe modulated precession of small arrays of vortices and antivortices. We place our findings in the context of similar results for other spatially confined nonlinear Hamiltonian systems in recent literature.

\end{abstract}

\maketitle

\section{Introduction}

Rich and complex dynamical phenomena often emerge when nonlinear waves are subject to spatial confinement. In the absence of dispersal to infinity, wave interactions are reinforced for an unlimited duration of time, giving the system ample opportunities to develop sophisticated traits in its evolution. Such features are especially pronounced for systems with fully resonant linearized spectra of frequencies. In this case, resonant interaction may generate elaborate dynamical phenomena for arbitrarily small nonlinearities, provided that one waits long enough.

The two-dimensional Gross-Pitaevskii equation for a Bose-Einstein condensate in a harmonic trap is an exemplary representative of the type of nonlinear dynamics we have just described. While it is of real-world significance as a model of effective dynamics of cold atomic gases \cite{BDZ,cooper,fetter}, it is also fascinating from a purely mathematical perspective for its combination of phenomenological complexity and rigid algebraic structure.

The linearized spectrum of the Gross-Pitaevskii equation is simply the perfectly resonant evenly-spaced energy spectrum of the harmonic oscillator, ensuring that significant nonlinear effects survive down to arbitrarily small values of the coupling parameter. Focusing on the small coupling asymptotics, one applies the time averaging method, which generates the corresponding {\em resonant system} \cite{GHT,BBCE}. The resonant system is a simplified infinite-dimensional Hamiltonian system accurately describing the original equation in the weakly nonlinear regime. It possesses extra structure and extra conserved quantities relative to the Gross-Pitaevskii equation.

In our previous work \cite{BBCE}, we focused on the consistent trucation of the resonant Gross-Pitaevskii system to the lowest Landau level which comprises only modes with maximal angular momenum at each energy level. We showed that the resulting lowest Landau level equation \cite{GHT,GT,GGT} admits explicit solutions describing a modulated precession of a single vortex around the center of the harmonic trap. Our purpose in the present article is to demonstrate that a similar picture emerges for two other types of consistent truncations: first, to sets of modes of fixed angular momentum, and second, to excited Landau levels.

The solutions we present are part of a bigger story that has emerged in recent studies of weakly nonlinear dynamics of resonant PDEs. The resonant system of the Gross-Pitaevskii equation shares its structure \cite{BMP} with various resonant systems emerging in Anti-de Sitter (AdS) spacetime \cite{FPU,CEV1,CEV2,CF,BEL}, in particular those studied in relation to the AdS stability problem \cite{BR,BMR,rev2}. This is no coincidence as the Gross-Pitaevskii equation can be seen as a non-relativistic limit of AdS wave equations \cite{BEL}. Some of the resonant systems we have mentioned admit special solutions very similar to the ones we shall present in this paper, and to the ones presented in \cite{BBCE}. This alludes to a common underlying pattern that shall be described elsewhere \cite{AO}. We also mention the cubic Szeg\H o equation introduced in \cite{GG} as an integrable model of turbulent energy transfer. This equation is algebraically very similar to the class of resonant systems that we focus on and possesses extremely rich dynamics which is analytically tractable due to its complete integrability.

We shall now proceed with our main exposition, split into three sections. The first section reviews the Gross-Pitaevskii equation and the construction of its resonant system, first studied in detail in \cite{GHT}, while the second and third sections present the solutions in the fixed angular momentum and excited Landau level truncations, which are the main target of our work. Important identities involving Laguerre polynomials and crucial for the construction of our solutions are given in appendices.

\section{The Gross-Pitaevskii\\ equation and its resonant system}

The Gross-Pitaevskii equation
\begin{equation}
i \del_t\Psi =\frac12\left(-\del_x^2-\del_y^2+x^2+y^2\right)\Psi + g|\Psi|^2\Psi
\label{GP}
\end{equation}
describes the evolution of the Bose-Einstein condensate wavefunction $\Psi(t,x,y)$ subject to an external harmonic potential $(x^2+y^2)/2$ referred to as the `trap.' (It is understood that in real-world implementations the condensate is narrowly confined in the remaining $z$-direction.)
The condensate self-interaction is characterized by the dimensionless coupling constant $g$.
We shall focus on studying this equation at small values of the coupling $0<g\ll 1$ (the sign of the coupling in fact does not affect the weakly nonlinear dynamics).

Equation (\ref{GP}) enjoys a large group of symmetries, known as the Schr\"odinger group \cite{Niederer,OFN}, which is in fact isomorphic to the symmetry group in the absense of any potential. This group includes boost-like generators (an analog of Galilean boosts) that permit, for example, boosting a static solution along classical harmonic oscillator trajectories \cite{translations}. We shall see below that extra symmetries emerge in the weakly nonlinear limit.

The linearized problem ($g=0$) corresponding to (\ref{GP}) is an ordinary two-dimensional isotropic harmonic oscillator with eigenvalues $E_n=n+1$ and normalized eigenfunctions (see, e.g., \cite{dahl}) given by
\beq\label{Psimn}
\Psi_{nm}\hspace{-1mm}=
\sqrt{\frac1\pi\frac{((n-|m|)/2)!}{((n+|m|)/2)!}}\,
r^{|m|}L^{\sssty|m|}_{\frac{\sssty n\hspace{-1pt}-\hspace{-1pt}|\hspace{-0.5pt}m\hspace{-0.5pt}|}2}(r^2)\,e^{-r^2/2}e^{im\phi}.
\eeq
Here, the energy level index $n$ is a nonnegative integer and the angular momentum index $m\in\{-n,-n+2,\dots,n-2,n\}$. The generalized Laguerre polynomials $L^\alpha_n(x)$ can be defined through the generating function
\beq\label{genLaguerre}
G_\alpha(t,x)=\sum_{n=0}^\infty t^n L^\alpha_n(x)=\frac{e^{-\frac{tx}{1-t}}}{(1-t)^{\alpha+1}}.
\eeq
They are orthogonal with the weight $x^{\alpha}e^{-x}$ on the interval $0\leq x< \infty$.

As remarked in the introduction, because the linearized solutions oscillate with integer frequencies, there are many resonances and effects of nonlinearities survive down to arbitrarily small coupling values. In such situations, naive expansion of solutions in powers of $g$ leads to growing `secular' terms that invalidate the naive perturbative expansion at times of order $1/g$. In order to capture the interesting weakly nonlinear dynamics, which happens precisely on time scales of order $1/g$, an alternative treatment is needed, and the time-averaging method \cite{murdock} provides a convenient framework. We shall only present a quick practical summary of the time-averaging here, referring the reader to \cite{murdock} for justification and proofs. One starts by expanding the exact solution $\Psi$ in terms of the linearized solutions
\begin{equation}
\Psi(t,r,\phi) = \sum_{nm} \alpha_{nm} (t)\,e^{-iE_nt} \Psi_{mn}(r,\phi).
\label{expand}
\end{equation}
 Substituting (\ref{expand}) to (\ref{GP}), one gets
\begin{equation}
i\,\frac{d\alpha_{nm}}{dt}\!=\! g\hspace{-8.5mm}  \sum_{\begin{array}{c}\ssty n_1,n_2,n_3\geq 0\vspace{-1.5mm}\\\ssty m+m_1=m_2+m_3\end{array}}\hspace{-7mm}\!
C_{n n_1 n_2 n_3}^{m m_1 m_2 m_3} \bar\alpha_{n_1m_1}\alpha_{n_2m_2}\alpha_{n_3m_3} e^{-i E t},
\label{GP_beta}
\end{equation}
where $E=E_n+E_{n_1}-E_{n_2}-E_{n_3}$ and we have introduced
\beq
C_{n n_1 n_2 n_3}^{m m_1 m_2 m_3}=\int \Psi_{nm} \Psi_{n_1m_1}\Psi_{n_2m_2}\Psi_{n_3m_3} \,r\,dr\,d\phi,
\eeq
which quantify the mode couplings and shall be called the interaction coefficients. The terms with $E=0$ correspond to resonant interactions while those with $E\neq 0$ are non-resonant.
Time-averaging amounts to introducing the \emph{slow time} $\tau=g t$ and discarding in \eqref{GP_beta}  all non-resonant terms, which oscillate rapidly in terms of $\tau$. The resulting equation (called the time-averaged or the resonant system) takes the  form
\begin{equation}
i\,\dot\alpha_{nm}=\hspace{-7mm}\sum_{\begin{array}{c}\ssty n+n_1=n_2+n_3\vspace{-1.5mm}\\\ssty m+m_1=m_2+m_3\end{array}}\hspace{-6mm}\!
C_{n n_1 n_2 n_3}^{m m_1 m_2 m_3}\bar\alpha_{n_1m_1}\alpha_{n_2m_2}\alpha_{n_3m_3},
\label{resGP}
\end{equation}
where from now on an overdot denotes $d/d\tau$.

Standard mathematical results on time-averaging \cite{murdock} guarantee that solutions of (\ref{resGP}) approximate solutions of (\ref{GP}) with arbitrarily high precision on time scales of order $1/g$ for sufficiently small coupling. (By contrast, on longer time scales, for example $1/g^2$, the two equations do not have to agree, and this is what leaves room for simplifications in the time-averaged system relative to the original equations.) While textbook discussions such as \cite{murdock} typically phrase their proofs in the context of finite-dimensional systems, a mathematical analysis of the validity of time-averaging specifically adapted to nonlinear Schr\"odinger equations (of which the Gross-Pitaevskii equation is an example) can be found in \cite{KM}.
General properties of the resonant system (\ref{resGP}) have been analyzed in \cite{GHT}. In particular, it was shown in \cite{GHT}  that (\ref{resGP}) enjoys many conservation laws, some of which  have no  counterparts for the original equation (\ref{GP}), and hence they are only approximately conserved by (\ref{GP}), though the precision of their conservation becomes arbitrarily good for small coupling. We shall see below the restrictions imposed by these general conservation laws on the specific truncations of (\ref{resGP}) we are interested in.

We note that because the $n$- and $m$-conservation conditions present in the sum in (\ref{resGP}), the equation can be consistently truncated to any set of modes satisfying $An+Bm=C$ with arbitrary numbers $A,B,C$. If only modes satisfying this relation are nonzero in the initial state, no other modes will get excited in the course of evolution. We are specifically concerned with two types of such truncations. First, we can retain only modes of some fixed angular momentum $m=\mm$, which without loss of generality we shall assume to be nonnegative. Second, we can retain modes with $n-m=2L$, where $L$ is referred as the `Landau level' number. (The lowest Landau level has previously received a good amount of attention, including treatments along our present lines in \cite{BBCE,GGT}.) The two types of truncations are depicted in fig.~1.
 \begin{figure}[t]
  	\begin{center}
  		\includegraphics[width=\columnwidth]{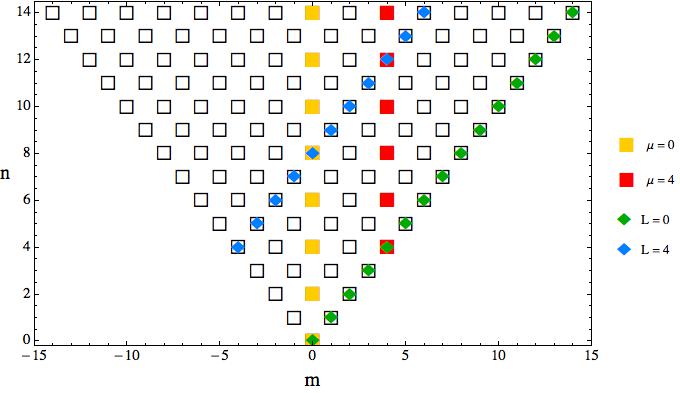}\vspace{-5mm}
  	\end{center}
  	\caption{\small{Truncations of the tower of modes of the two-dimensional Gross-Pitaevskii equation to fixed angular momentum (vertical groups of modes) and Landau levels (diagonal groups of modes).}}\vspace{-2mm}
  	\label{fig1}
  \end{figure}

\section{Solutions for angular momentum truncations}

To deal with the resonant system (\ref{resGP}) truncated to modes of angular momentum $\mu$, we introduce the following notation for the relevant modes:
\beq
\beta_n\equiv \sqrt\frac{n!}{(n+\mm)!}\,\, \alpha_{\mm+2n,\mm},
\eeq
while setting all other modes to zero. 
With these specifications, (\ref{resGP}) reduces to
\beq\label{resbeta}
i \,\frac{(n+\mm)!}{n!}\, \dot \beta_{n}=\sum_{j=0}^{\infty}\sum_{k=0}^{n+j}S_{njk,n+j-k}\bar\beta_{j}\beta_{k}\beta_{n+j-k},
\eeq
where
\beq\label{defS}
{S}_{njkl}= \int\limits_0^\infty d\rho \, e^{-2\rho} \rho^{2\mm} \,L^{\mm}_{n}(\rho) L^{\mm}_{j}(\rho) L^{\mm}_{l}(\rho) L^{\mm}_{k}(\rho).
\eeq
(Throughout the article, we ignore in (\ref{resbeta}) and further equations derived from it purely numerical factors that can be absorbed in a redefinition of time.)

The integral in (\ref{defS}) can in principle be evaluated by substituting explicit expressions for the Laguerre polynomials given by
\beq
L_n^\mm(\rho)=\sum_{k=0}^n (-1)^k\frac{(n+\mm)!}{k!\,(n-k)!(k+\mm)!}\rho^k
\label{Lag_explicit}
\eeq
and evaluating the remaining integrals of the form $\int d\rho\,\rho^A e^{-2\rho}$. The resulting expression is, however, a quadruple sum with summation ranges depending on $n$, $j$, $k$ and $l$. We have not been able to bring this sum to a manageable form that could be employed in explicit derivations. Instead, our subsequent analysis will rely directly on the integral representation (\ref{defS}) and identities satisfied by the Laguerre polynomials.

Note that the fixed angular momentum truncation to wavefunctions of the form $\Psi(r,t)e^{i\mm \phi}$ is equally valid in the Gross-Pitaevskii equation (\ref{GP}). This is in contrast to the lowest Landau level truncation of \cite{BBCE}, and the excited Landau level truncations we shall consider below, which hold exactly only within the resonant system (\ref{resGP}). One simply substitutes $\Psi(r,t)e^{i\mm \phi}$ in (\ref{GP}), which results in a radial Gross-Pitaevskii equation with the effective potential $r^2+\mm^2/r^2$, sometimes referred to as the `pseudoharmonic oscillator' potential \cite{dong}. The two-dimensional Gross-Pitaevskii resonant system (\ref{resbeta}) can be alternatively viewed as the resonant system of this effective one-dimensional radial equation.  We remark parenthetically that all the formulas we present below hold for any real positive values of $\mm$ (even if only integer values arise starting from the two-dimensional Gross-Pitaevskii equation), provided that factorials $x!$ are replaced by the gamma functions $\Gamma(x+1)$.

As follows from general considerations of \cite{GHT}, the resonant system (\ref{resbeta}) respects conservation of the following quantities: the particle number
\beq
N=\sum_{n=0}^\infty \frac{(n+\mm)!}{\mm!n!} |\beta_n|^2,
\eeq
the `linear energy'
\beq
E=\sum_{n=0}^\infty (n+1)\frac{(n+\mm)!}{\mm!n!} |\beta_{n}|^2,
\eeq
the Hamiltonian
\beq
H=\sum_{n,j=0}^\infty\sum_{k=0}^{n+j}S_{njk,n+j-k}\bar\beta_{n}\bar\beta_{j}\beta_{k}\beta_{n+j-k},
\eeq
and
\beq\label{Z}
Z=\sum_{n=0}^\infty \frac{(n+1+\mm)!}{(\mm+1)!n!}  \bar\beta_{n+1} \beta_{n}.
\eeq

Motivated by considerations of similar systems in \cite{BBCE,CF,BEL}, we shall now look for special solutions of (\ref{resbeta}-\ref{defS}) in the form
\beq\label{3dansatz}
\beta_n(\tau)=\left(b(\tau)+\frac{a(\tau)n}{p(\tau)}\right)(p(\tau))^n,
\eeq
where $a$, $b$ and $p$ are complex-valued functions of time.
While it is by no means obvious that this ansatz provides a consistent truncation of (\ref{resbeta}-\ref{defS}), this can be demonstrated with the use of the following identities:
\begin{align}
&\sum_{k=0}^{n+j} S_{njk,n+j-k} = \frac{(2\mm)!}{2^{2\mm+1}} \frac{(n+\mm)!}{\mm! n!}\frac{(j+\mm)!}{\mm! j!},\label{lagid1}\\
&\sum_{k=0}^{n+j} k S_{njk,n+j-k}=  \frac{(2\mm)!}{2^{2\mm+1}} \frac{(n+\mm)!(j+\mm)!}{(\mm!)^2n!j!} \frac{(n+j)}2,\label{lagid2}\\
&\sum_{k=0}^{n+j}k(n+j-k)S_{njk,n+j-k}= \frac{(2\mm)!}{2^{2\mm+1}} \frac{(n+\mm)!(j+\mm)!}{(\mm!)^2n!j!}\nonumber \\
&\hspace{5mm}\times\left[\frac{(n+j)(n+j-1)}8+\frac{1+2\mm}{4(1+\mm)}nj\right].\label{lagid3}
\end{align}
Proofs of these identities are given in appendix A.

Substituting (\ref{3dansatz}) in (\ref{resGP}) and using the above summation identities results in a statement that two quadratic polynomials in $n$ equal each other. Equating the coefficients of these polynomials results in three first order ordinary differential equations for the three functions $a(\tau)$, $b(\tau)$ and $p(\tau)$:
\onecolumngrid
\begin{align}
 \frac{1}{(1+\mm)}\frac{8i\dot{b}}{(1+y)^{2+\mm}} =& 8 b \Big[\frac{|b|^2}{(1+\mm)(1+y)}
+ (1+(2+\mm)y) |a|^2  +\bar{a} b p + a \bar{b}\bar{p}\Big]\label{eom_b} \\
& \hspace{3cm}+ a \bar{p} (2+\mm) (1+y) \left[a \bar{b}\bar{p} +  (2 + (3 + \mm) y) |a|^2\right],\nonumber \\
 \frac{8i\dot{a}}{(1+y)^{2+\mm}} =&a \Big[(4+6\mm) a \bar{b}\bar{p} + 7(1+\mm) \bar{a} b p
 + (4+6\mm)(1 + (2+\mm) y) |a|^2+  \frac{7 |b|^2}{1+y}\Big]\label{eom_a}, \\
 \frac{8i\dot{p}}{(1+y)^{2+\mm}} =&a \left(\frac{\bar{b}}{1+y} + (1 + \mm) \bar{a} p\right),\label{eom_p}
\end{align}
where we have rescaled the time $\tau \rightarrow (2\mm)!/(2^{2\mm+1}\mm!)\tau$ and introduced an auxiliary quantity
\beq
y = \dfrac{|p|^2}{1-|p|^2}.
\eeq
Within the ansatz (\ref{3dansatz}) the conserved quantities take the form:
\begin{align}
& \frac{N}{(1+\mm)(1+y)^{2+\mm}} =
  (a \bar{b} \bar{p}+ \bar{a}b p) +  (1 + (\mm+2) y)|a|^2 + \frac{|b|^2}{(1+\mm)(1+y)},\label{cc_N}\\
&  \frac{E}{(1+\mm)(1+y)^{2+\mm}} =
   \frac{1+ (1+\mm) y}{ (1+\mm) (1+y)} |b|^2+ (2 + (2+\mm) y)\left(\bar{a} b p + a \bar{b} \bar{p}\right)  +  (2 +  (8+4\mm) y + (6 + 5 \mm + \mm^2) y^2)  |a|^2,\label{cc_E}\\
& \frac{Z}{(2+\mm)(1+y)^{\mm+3}} =
\bar{p} \left( \frac{|b|^2}{(2+\mm)(1+y) } + a \bar{b}\bar{p} + (2+(3+\mm)y)|a|^2\right)
 +\frac{1+(2+\mm)y}{(2+\mm)(1+y)} \bar{a}b. \label{cc_Z}
\end{align}
\hrulefill
\twocolumngrid
\noindent The Hamiltonian can be expressed as
\beq\label{H_ansatz}
H = \frac{(2\mm)!}{2^{2\mm+1}(\mm!)^2}\left( N^2 - \frac{2+\mm}{4(1+\mm)}S^2 \right),
\eeq
where
\beq\label{cc_S}
S =  (1+y)^{3+\mm} (1+\mm) |a|^2.
\eeq
Using the above conservation laws, one can simplify the system (\ref{eom_b}-\ref{eom_p}) to the form
\begin{align}
& 8i\dot{b} = 8 N b + (2+\mm)\big[\bar{p}(1+y)(E-N+S) \label{bs} \\&\hspace{3.5cm} - (1+\mm)yZ\big] a, \nonumber\\
& 8(1+\mu)i\dot{a} =  \big[(10+8\mm) N -(3+\mm) E\,\label{as} \\
& \hspace{3cm}+(1+\mm)(3+\mm) p Z\big] a, \nonumber\\
& \hspace{7mm}8 (1+y) i \dot p = \bar Z - \frac{E+S+\mm N}{1+\mm}\, p. \label{ps}
\end{align}\
This system can be integrated by first solving equation \eqref{ps}  and then substituting  $p(\tau)$  into equations \eqref{as} and \eqref{bs}, which yields two linear equations for $a(\tau)$ and $b(\tau)$. Here we write only the solution for $y(\tau)$, which is remarkably simple. It follows from \eqref{ps} that
\beq\label{eq_yp2}
\dot y^2 =\frac{1}{16} (1+y)^2 \left[\Im(Zp)\right]^2.
\eeq
Using the conservation laws, we find that the right hand side of \eqref{eq_yp2} is a quadratic function of $y$ with coefficients depending of $N, E$, and $S$. The equation thus looks like energy conservation of a harmonic oscillator in the $y$-direction, and is integrated as
\beq\label{sol_y}
y(\tau) = A\sin(\Omega \tau + \theta) + B,
\eeq
with
\begin{align}
& A = \frac{\sqrt{S(N-S) ((E-N) (E+\mm N) - (2+\mm) S^2) }}{32(1+\mu)^{3/2}\Omega^2} \label{A_sol}\\
& B =  \frac{(1+\mm) N (E-N+S)+ 2 S (E-N-(2+\mm) S)}{64 (1+\mm)^2 \Omega^2}\label{B_sol}\\
& \Omega = \frac{1}{8}\sqrt{N^2 + \frac{8+4\mm}{(1+\mm)^2}S^2}\,. \label{Omega_sol}
\end{align}
Through the conservation laws \eqref{cc_N}, \eqref{cc_E}, and \eqref{cc_S}, the periodic behavior of $y(t)$ is transferred to $|a(t)|^2$, $|b(t)|^2$, and $\Re(\bar a b p)$, and hence to the mode spectrum
\beq\label{energy_spectrum}
|\beta_n|^2=  \left(|b|^2+\frac{2\Re(\bar a b p)}{|p|^2}\,n +\frac{|a|^2}{|p|^2}\,n^2\right) |p|^{2n}\,.
\eeq
The turning points of the oscillations of $y(t)$, given by $y_{\pm} = B \pm A$,  provide lower and upper bounds for the inverse and direct cascades of energy, respectively.
 Using \eqref{A_sol}-\eqref{B_sol}, and following the strategy of \cite{CF}
we obtain a rough bound
\beq\label{ypm_bound}
\frac{1+y_{+}}{1+y_{-}}\leq \left(\frac{6+2\mm}{1+\mm}\right)^2,
\eeq
which proves that the transfer of energy to high modes is uniformly bounded.

The special case $A=0$ corresponds to stationary solutions for which $|\beta_n|^2$ are constant. For such solutions,
\begin{equation}\label{stationary}
  \beta_n(\tau)=A_n\,e^{-i (\lambda-\omega n) \tau},
\end{equation}
where the amplitudes $A_n$ are time-independent and the parameters $\lambda$ and $\omega$ are real-valued.
Within the ansatz \eqref{3dansatz} they take the form
\begin{equation}\label{stationary_3d}
  b(\tau)=\beta e^{-i\lambda \tau},\, a(\tau)=\gamma e^{-i(\lambda-\omega)\tau},\, p(\tau)=q e^{i\omega \tau}.
\end{equation}
Thanks to the phase rotation symmetries of the resonant system~\eqref{resbeta} $\beta_n \mapsto e^{i\theta} \beta_n$ and $\beta_n \mapsto e^{i n \theta} \beta_n$,  one can assume without loss of generality that the parameters $\beta, \gamma, q$ are real-valued. Substituting \eqref{stationary_3d} into the system (\ref{eom_b}-\ref{eom_p}),  we get an algebraic system which has four two-parameter families of solutions:
\begin{enumerate}
\item[(a)] $\begin{aligned}[t]\label{a}
 \beta&=c,\quad \gamma=0, \quad \omega=0,\quad \lambda=\frac{c^2}{(1-q^2)^{1+\mm}}\,;
\end{aligned}$
\item[(b)] $\begin{aligned}[t]\label{b}
   \beta&=c\, q,\quad \gamma=-c\, \frac{1-q^2}{1+\mm}, \quad \omega=0,\\
   \lambda&=\frac{c^2}{4 (1-q^2)^{1+\mm}} \frac{2+3\mm}{(1+\mm)^2}\,;
\end{aligned}$\vspace{2mm}
\item[(c$_\pm$)] \hspace{-1mm}$\begin{aligned}[t]\label{c}
   &\beta_{\pm}=\frac{c}{2}\,\left(1+(3+2\mm) q^2 \pm \kappa\right), \\ 
   &\gamma=-c\, q (1-q^2),\\
  &\hspace{-6mm} \lambda_{\pm}=\frac{c^2}{16 (1-q^2)^{\mm}} \!\left(8\!-\!(9+\mm) q^2 \!\pm \!\frac{8+(9+\mm) q^2}{1-q^2}\kappa\!\right)\\
   &\omega_{\pm}=\frac{c^2}{16 (1-q^2)^{\mm}}\,(1+q^2\pm \kappa),
   \end{aligned}$
\end{enumerate}
where $\kappa=\sqrt{q^4-(10+4\mm)q^2+1}$ and $c$, $q$ are real-valued parameters. For the solutions (a) and (b) the range of $q$ is $0\leq q<1$, while for the solutions (c) $0\leq q^2<2\mm+5-2\sqrt{\mm^2+5\mm+6}$.
The stationary states (a) and (c$_+$) bifurcate at $q=0$ from the $n=0$ single-mode state, while the states (b) and (c$_-$) bifurcate from the $n=1$ single-mode state.

We remark that the families of stationary solutions (a) and (b) are nothing else but the symmetry orbits of the single-mode states $\beta^{(0)}_n=\delta_{n0} e^{-it}$ and $\beta^{(1)}_n=\delta_{n1} e^{-\frac{i}{2} (2+3\mu)t}$, respectively,
generated by the symmetry $\beta_n \mapsto e^{\xi D} \beta_n$, where $D$ is the Hamiltonian vector field associated with the conserved quantity $\Im{Z}$, given by
\begin{equation}
  D \beta_n :=\{\Im{Z},\beta_n\}= n\beta_{n-1} - (n+\mu+1) \beta_{n+1}.
\end{equation}
(The Poisson brackets are evaluated with respect to the symplectic form $i\sum_n d\bar\beta_n\wedge d\beta_n (n+\mu)!/n!$.)
On the three-dimensional invariant subspace \eqref{3dansatz} this symmetry can be shown to act as follows
\begin{align}
& p\mapsto \frac{p+\tanh{\xi}}{1+p \tanh{\xi} }, \quad a\mapsto \frac{a}{(\cosh{\xi}+p \sinh{\xi} )^{3+\mu}},\nonumber\\
& b \mapsto \frac{b(1+p \tanh{\xi})-a(1+\mu) \tanh{\xi}}{(1+p \tanh{\xi})(\cosh{\xi}+p \sinh{\xi})^{1+\mu}} .
\end{align}
Applying this transformation to the single-mode states $\beta^{(0)}$ (for which $b=e^{-i\tau}, a=0, p=0$)
and $\beta^{(1)}$ (for which $b=0, a=e^{-\frac{i}{2}(2+3\mu)\tau}, p=0$), we obtain (modulo rescaling) the stationary states (a) and (b) with $q=\tanh{\xi}$, respectively.

Transformations generated by $N$, $E$, $Z$ can in fact be used to convert the stationary states (c$_{\pm}$) (for which $Z=0$) into dynamical solutions within the ansatz (\ref{3dansatz}). Indeed, solutions (c$_{\pm}$) have two parameters $c$ and $q$, quantifying the overall scale of $a$ and $b$, and the absolute value of $p$, respectively. By acting with the above transformations generated by $\Im{Z}$ and the following transformations generated by $\Re{Z}$,
\begin{align}
& p\mapsto \frac{p - i\tanh\eta}{1 +i p \tanh\eta}, \quad 
a\mapsto \frac{a}{(\cosh\eta + ip \sinh\eta)^{3+\mu}},\nonumber\\
& b \mapsto  \frac{ b(1 +i p \tanh\eta) -i (1+\mm) a  \tanh\eta}{(1 +i p \tanh\eta)(\cosh\eta + i p \sinh\eta)^{1+\mm}} ,
\end{align}
one can adjust the magnitude and phase of the ratio $a/b$. Thereafter, the
phase rotation symmetries $\beta_n \mapsto e^{i\theta} \beta_n$ and $\beta_n \mapsto e^{i n \theta} \beta_n$ (generated by $N$ and $E$) can be used to freely adjust the overall phase of $b+an/p$ and, independently, the phase of $p$. This construction of dynamical solutions
naturally explains the periodicity of $y$ we observed by solving (\ref{eq_yp2}) directly through the simple periodicity of $p$ in
the stationary states (c$_\pm$).

The state (a) is the ground state in the sense that it is a maximizer of $H$ for fixed $N$, which follows within the ansatz \eqref{3dansatz} from \eqref{H_ansatz}. Hence it is orbitally stable \cite{GHT} (cf.~\cite{BHP1} for the proof of the analogous result for the cubic conformal flow on $S^3$). The question of stability of the stationary states (b) and (c$_{\pm}$) is open. Another interesting open problem is the classification of all stationary states of the resonant system \eqref{resbeta} (see \cite{GGT} and \cite{BHP2} for analysis of the corresponding problem for the LLL equation and the cubic conformal flow on $S^3$, respectively).


\section{Solutions for Landau level truncations}

Truncation of (\ref{resGP}) to the excited Landau levels $n-m=2L$ is in principle completely straightforward. However, the explicit form of linear wavefunctions (\ref{Psimn}) is rather inconvenient for this purpose, since both positive and negative values of $m$ are present within each excited Landau level (see fig.~1) and hence the absolute values of $m$ present in (\ref{Psimn}) make dependences on index numbers rather awkward. To remedy this unwelcome feature, we shall start by rewriting (\ref{Psimn}) in a slightly different form using identities satisfied by the Laguerre polynomials.

From (\ref{Lag_explicit}), remembering that factorials of negative numbers are infinite, one gets
\begin{equation}
L_{n}^{\alpha}(\rho) = (-1)^{\alpha}\frac{(n+\alpha)!}{n!}\rho^{-\alpha}L_{\alpha+n}^{-\alpha}(\rho)
\end{equation}
for any integer $\alpha$. (Note that $L^{-\alpha}_{\alpha+n}$ does not contain any powers of $\rho$ below $\rho^\alpha$.)
Using this formula, we obtain
\begin{align}
&\sqrt{\frac{((n-|m|)/2)!}{((n+|m|)/2)!}}\rho^{|m|/2}L_{\frac{n-|m|}{2}}^{|m|}(\rho) \label{laginvert}\\
&\hspace{5mm}= (-1)^{\frac{1}{2}(m - |m|)} \sqrt{\frac{((n-m)/2)!}{((n+m)/2)!}}\rho^{m/2}L_{\frac{n-m}{2}}^{m}(\rho).\nonumber
\end{align}

We then define
\beq
\beta_n=(-1)^{\frac{|n-L|-(n-L)}{2}}\alpha_{n+L,n-L},
\label{betaLL}
\eeq
with $n$ running from 0 to $\infty$.
The sign factor inserted compensates for the sign factor in (\ref{laginvert}), simplifies the subsequent expressions, and brings our sign conventions in accord with \cite{GHT} for the remainder of our treatment. Truncating (\ref{resGP}) to the above set of modes results in
\beq
i \,\dot \beta_{n}=\sum_{m=0}^{\infty}\sum_{k=0}^{n+m}S_{nmk,n+m-k}\bar\beta_{m}\beta_{k}\beta_{n+m-k},
\label{LLflow}
\eeq
with
\begin{align}
&\tildeC_{nmkl} = \int\limits_{0}^{\infty}d\rho\, e^{-2\rho} \,\frac{L!^2 \rho^{n+m-2L}}{\sqrt{n!m!k!l!}} \label{S_LL}\\
&\hspace{15mm}\times L_{L}^{n-L}(\rho)L_{L}^{m-L}(\rho)L_{L}^{k-L}(\rho)L_{L}^{l-L}(\rho).\nonumber
\end{align}
We caution the reader that the letters $\beta$ and $S$ in this and the previous section refer to similar quantities in two different truncations of (\ref{resGP}) and should not be identified. As before, $S$ can be computed explicitly using (\ref{Lag_explicit}), but such expressions contain multiple sums and are of little practical use. Our derivations will rely on identities satisfied by the Laguerre polynomials, and not on awkward explicit expressions for $S$. We note the following simple generating function, derived explicitly in appendix B, for the sequence of polynomials appearing in (\ref{S_LL}):
\beq
G_L(s,\rho)=\sum_{n=0}^\infty \frac{s^n}{n!} L_L^{n-L}(\rho)
=\frac{e^s(s-\rho)^L}{L!}.\label{G_L}
\eeq

Equation (\ref{LLflow}) respects the following conserved quantities, in agreement with \cite{GHT}:
\begin{align}
&N = \sum_{n=0}^{\infty}|\beta_n|^{2},\\ 
&J = \sum_{n=1}^{\infty} n |\beta_n|^{2},\\
&H=\sum_{n,m=0}^\infty\sum_{k=0}^{n+m}S_{nmk,n+m-k}\bar\beta_{n}\bar\beta_{m}\beta_{k}\beta_{n+m-k},\\
&Z = \sum_{n=0}^{\infty} \sqrt{n+1}\,\bar{\beta}_{n+1} \beta_n.
\end{align}

It turns out that (\ref{LLflow}) is rather similar to the lowest Landau level equation \cite{GHT,BBCE,GGT} and admits solutions of the form
\begin{equation}\label{eq:ansatz_L_1}
\beta_n = \frac{1}{\sqrt{n!}}\left(b(\tau) + \frac{a(\tau)}{p(\tau)} n\right)(p(\tau))^n.
\end{equation}
The closure of the ansatz relies on the following three summation identities, which we prove in appendix B:
\begin{align}
&\sum_{k=0}^{n+m} \sqrt{\frac{n!m!}{k!(n+m-k)!}}\tildeC_{nmk,n+m-k}=\frac{(2L)!}{2^{2L+1}(L!)^2}.\label{LLid1}\\
&\sum_{k=0}^{n+m} k\sqrt{\frac{n!m!}{k!(n+m-k)!}}\tildeC_{nmk,n+m-k}\label{LLid2}\\
&\hspace{3cm}=\frac{(2L)!}{2^{2L+1}(L!)^2}\frac{n+m}{2}.\nonumber\\
&\sum_{k=0}^{n+m} k(n+m-k) \sqrt{\frac{n!m!}{k!(n+m-k)!}}\tildeC_{nmk,n+m-k}\nonumber\\
&\hspace{7mm}=\frac{(2L)!}{2^{2L+1}(L!)^2}\Bigg[\frac{L-1}{4(2L-1)}(n+m)(n+m-1)\nonumber\\
&\hspace{4cm}+\frac{L}{2L-1}nm\Bigg].\label{LLid3}
\end{align}
Substituting (\ref{eq:ansatz_L_1}) in (\ref{LLflow}) results (in a repetition of a pattern encountered in the previous section) in a statement that two quadratic polynomials in $n$ equal each other. Equating the three coefficients of these polynomials results in three first order ordinary differential equations for $a$, $b$, $p$.

It is convenient to present the equations for $a$, $b$, $p$ in a compact form that takes into account the conservation laws. Within the ansatz (\ref{eq:ansatz_L_1}), the conserved quantities are:
\begin{align}
&N = e^{|p|^2}\left(|b|^2 + (1+|p|^2)|a|^2 + (a\bar{b}\bar{p} + \bar{a} b p)\right),\label{LL_N}\\
&J = e^{|p|^2}\Big(|p|^2 |b|^2 + (1+3|p|^2 + |p|^4)|a|^2 \label{LL_J}\\
&\hspace{3cm}+ (1+|p|^2)(a\bar{b}\bar{p} + \bar{a} b p)\Big),\nonumber\\
&Z = e^{|p|^2}\Big(\bar{p} (|b|^2 + (2+|p|^2) |a|^2 + a \bar{b}\bar{p})\label{LL_Z}\\
&\hspace{4cm} + (1+|p|^2)\bar{a}b \Big).\nonumber
\end{align}

There are two distinct cases to consider. First, for $L=1$, upon time rescaling  $\tau \rightarrow \frac{(2L)!}{2^{2L}(L!)^2 (2L-1)}\tau$, the equations for $a$, $b$, $p$ are written in the form
\begin{align}
& \,\,\dot{p} = 0,\\
& i\dot{a} = - N a/2,\\
& i\dot{b} =  - N b/2.
\end{align}
All solutions of this system are stationary and of the form $\beta_n(\tau) = (b(0)+ a(0) n)(p(0))^{n} e^{i N \tau/2}$.

For $L>1$, we perform the time rescaling $\tau \rightarrow \frac{(2L)! (L-1)}{2^{2L}(L!)^2 (2L-1)}\tau$ to obtain
\begin{align}
& 8i\dot{p} = \bar{Z} - N p,\label{eqpLL}\\
& 8i\dot{a} = \left( Zp - J + \frac{7L-3}{L-1} N\right) a,\\
& 8i\dot{b} =  Z a + \left(Zp  - J + \frac{8L-4}{L-1} N\right) b.
\end{align}
The equation for $\dot{p}$ is independent of $L$, as are the conserved quantities (\ref{LL_N}-\ref{LL_Z}) within the ansatz (\ref{eq:ansatz_L_1}).
Hence, $|b|^2$, $|a|^2$, $Re(a\bar{b}\bar{p})$ and $p$ have exactly the same behavior for any $L>1$ as they do for the lowest Landau level $L=0$, and so does the spectrum $|\beta_n|^2$. Since solutions at the lowest Landau level have already been treated in detail in \cite{BBCE}, we shall not repeat the derivations here.

\section{Discussion}

 \begin{figure*}[t!]
\begin{center}
\includegraphics[width=14cm]{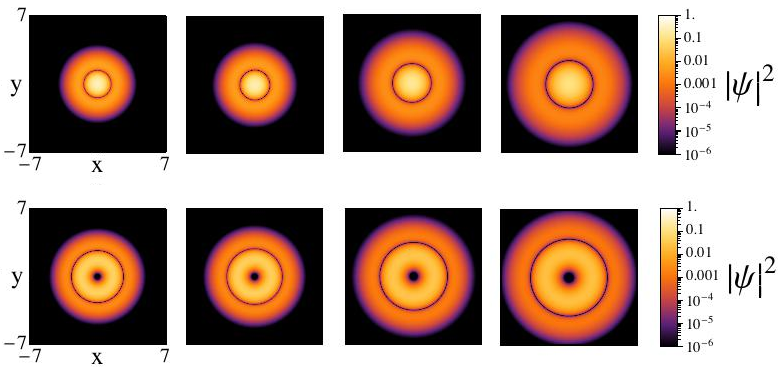}\vspace{-5mm}
\end{center} 
 	\caption{\small{Snaphots of the condensate density $|\psi(t,x,y)|^2$ for resonant solutions in the fixed angular momentum sector with $\mu=0$ (upper row) and $\mu=4$ (lower row). For both, the initial data used are a = 0.5, b = 0.5, p = 0, and the coupling is g = 0.01. The limits of the axes and their labels are identical to the first plot of each row. Time points are, from left to right, $t = 201,\ 3832.1,\ 7663.9,\ 11501.4$ for the upper row and $t = 219.9,\ 22003,\ 43999.4,\ 66001.7$ for the lower one.}}\vspace{-3mm}
  	\label{fig2}
  \end{figure*}
We have considered truncations of the Gross-Pitaevskii resonant system (\ref{resGP}) to sets of modes at fixed angular momentum and fixed Landau level (fig.~1) and demonstrated that such truncations admit special analytic solutions similar to the ones previously seen specifically for the lowest Landau level truncation \cite{BBCE}. The construction of our new solutions relies on sophisticated nonlinear identities satisfied by Laguerre polynomials and analyzed in the appendices. Similar solutions have recently appeared in the context of a number of other resonant systems \cite{CF,BEL}, while the general underlying theory has been developed concurrently with this article in \cite{AO}.

It is instructive to examine the position space form of our solutions. For fixed angular momentum truncations within the ansatz (\ref{3dansatz}), one finds (omitting an irrelevant overall phase factor)
\begin{align}
&\Psi(r,\phi,t)=\sum_n \sqrt\frac{(n+\mm)!}{n!} \beta_n e^{-i(\mm+2n)t}\Psi_{\mm+2n,\mm}(r,\phi)\nonumber\\
&\hspace{3mm}=r^\mm e^{i\mm(\phi-t)} e^{-r^2/2} \sum_n \left(b+\frac{a n}{p}\right)p^n e^{-2int}L^\mm_n(r^2) .
\end{align}
Using (\ref{genLaguerre}), one obtains
\begin{align}
\Psi&=r^\mm e^{i\mm(\phi-t)} e^{-r^2/2}\left(b+\frac{a}{p}\xi\del_\xi\right)G_\mm(\xi,r^2)\Big|_{\xi=s} \nonumber \\
&=\frac{r^\mm e^{i\mm(\phi-t)} e^{-\frac{r^2}{2}\frac{1+s}{1-s}}}{(1-s)^{\mm+1}}\,e^{-2it} \label{positionspace}\\
&\times\left(b(\tilde gt) e^{2it}+\frac{(\mm+1)a(\tilde gt)}{1-s}-\frac{a(\tilde gt) r^2}{(1-s)^2} \right),\nonumber
\end{align}
where $s(t)=p(\tilde gt)e^{-2it}$, and $a$, $b$, $p$ are functions of the slow time $\tilde gt$, as explicitly shown. (We have absorbed the $\mu$-dependent time rescaling factor mentioned under (\ref{eom_p}) into the modified coupling $\tilde g$.)
This represents rapidly oscillating rings of wavefunction density with the superposed slow periodic parameter modulations.

For nonzero angular momentum ($\mm>0$ in our conventions), the wavefunction vanishes at $r=0$ because of the usual angular momentum barrier. More interestingly, there is at most one more value of the radial coordinate where the wavefunction has a node; such a node appears if the expression in parentheses in (\ref{positionspace}) has a zero for positive $r^2$. Otherwise, there is not quite a node but just a dip in density. Snapshots of the condensate density of representative solutions can be seen on fig. 2. In the literature, such objects have been referred to as `ring dark solitons,' see e.g.\ \cite{Kevrekidis}. They have been mostly studied, however, in the presence of strong nonlinearity, where the solitons are supported by nonlinear effects and the harmonic trap is merely a perturbation. In contrast, in our weakly nonlinear regime, the `solitons' are mainly supported by the harmonic trap, with the nonlinearities giving rise to slow modulation of the parameters; see e.g.\ \cite{PK,Wang} for  discussions of other dark solitons near the linear regime.

Stability of dark rings with static ring profiles (i.e., time-independent $|\Psi|^2$) has been extensively studied in the literature on Bose-Einstein condensates \cite{ring1,ring2,ring3}. There are interesting instabilities trigerring pattern formation. While our solutions given by (\ref{positionspace}) also feature dark rings, the ring radius rapidly oscillates with the trap frequency, and there is no immediate mathematical relation to the dark rings with static profiles. Stability of our oscillating rings would be an interesting, if not obvious, question to investigate.  In \cite{PK}, the question of stability is treated for related oscillating dark solitons for the one-dimensional Gross-Pitaevskii equation in a harmonic trap. In \cite{Wang}, analogous oscillating dark shells in three dimensions are studied by predominantly numerical methods, with evidence for their stability at sufficiently small coupling.

Turning to the excited Landau levels, we use (\ref{Psimn}), (\ref{expand}), (\ref{laginvert}-\ref{betaLL}), (\ref{G_L}) and (\ref{eq:ansatz_L_1}) to find (omitting an irrelevant overall phase factor)
\beq
\Psi(t,r,\phi) = \sqrt{\frac{L!}{\pi}}\frac{e^{-r^2/2}}{z^L}
\left(b+ \frac{a}{p} \xi\partial_\xi\right) G_{L}(\xi, r^2)\big{|}_{\xi=pz},
\eeq
where $z \equiv r e^{i(\phi-t)}$. Direct evaluation yields
\interdisplaylinepenalty=10000
\begin{align}
&\Psi = \frac{e^{-r^2/2}e^{p(\tilde gt) z}}{\sqrt{L!\pi}}\left(p(\tilde gt)-\bar{z}\right)^{L-1}\label{PsiLL}\\
&\hspace{1cm}\times\left[(b(\tilde gt)+a(\tilde gt)z)(p(\tilde gt)-\bar{z})+a(\tilde gt)L\right],\nonumber
\end{align}
\interdisplaylinepenalty=0
where we have explicitly displayed the dependence of $a(\tilde gt)$, $b(\tilde gt)$, $p(\tilde gt)$ on the slow time $\tilde g t$. (Again,  the $L$-dependent time rescaling factor mentioned above (\ref{eqpLL}) has been absorbed by introducing $\tilde g$ instead of $g$.)
This expression describes periodically modulated (due to slow time dependence of $a$, $b$, $p$) precession around the origin of an array consisting of a degree $L-1$ antivortex at $z=\bar p$ and a vortex-antivortex pair whose location is given by the two zeros of the second line of (\ref{PsiLL}),
\beq
(z+b/a)(\bar z-p)=L.
\label{pairloc}
\eeq 
Snapshots of the corresponding wavefunction density are given in fig.~3. The three defects we mentioned (two antivortices and a vortex) in fact always lie on the same straight line. This can be seen by taking the imaginary part of (\ref{pairloc}),
\beq
\Im\left[z\left(p+\frac{\bar b}{\bar a}\right)+\frac{pb}a\right]=0,
\eeq
which defines a straight line to which all roots of (\ref{pairloc}) belong. Since $z=\bar p$ evidently satisfies the above equation, it lies on the same straight line.

Small arrays of vortices have been studied in the literature on Bose-Einstein condensates with variational methods \cite{arr1,arr2,arr3}, and have even been created experimentally \cite{arrexp}. In our context, special configurations in this class can be accessed by rigorous asymptotic methods, and we provide exact analytic solutions for the vortex dynamics. A significant improvement is thus achieved over the conventional variational techniques.

\onecolumngrid

\section{Acknowledgments}
This research has been supported by  CUniverse research promotion project (CUAASC), FWO-Vlaanderen (projects G044016N and G006918N), Polish National Science Centre grant no.\ 2017/26/A/ST2/00530, by Vrije Universiteit Brussel through the Strategic Research Program ``High-Energy Physics,'' by FPA2014-52218-P~from Ministerio de Economia y Competitividad, by Xunta de Galicia ED431C 2017/07, by the European Regional Development Fund (FEDER) and by Grant Mar\'\i a de Maeztu Unit of Excellence MDM-2016-0692. 

A.B. thanks Javier Mas and the support of the Spanish program ``Ayudas para contratos predoctorales para la formaci\'on de doctores 2015'' associated to FPA2014-52218-P and its mobility program for his stay at Jagiellonian University, where part of this project was developed.

\appendix

\section{Appendix A: Identities for the interaction coefficients of angular momentum truncations}

 \begin{figure*}[t!]
\includegraphics[width=14cm]{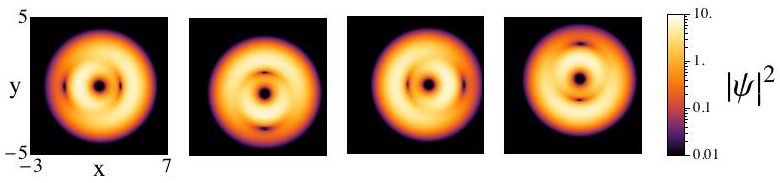}
  	\caption{\small{Snaphots of the condensate density $|\psi(\tau,z)|^2$ for resonant solutions in the excited Landau level sector at $L=4$. The representation is in the maximally rotating frame and the visible movements represent slow modulations described by our resonant solutions. In the laboratory frame, the whole picture rapidly rotates around the origin. The initial data used are a = 0.5, b = 0.5, p = 2, and the time points are $\tau = 0,\ 7.8,\ 15.6,\ 23.4$. The limits of the axes and their labels are identical to the first plot of the sequence.}}
  	\label{fig3}
  \end{figure*}
We shall now demonstrate how one can derive (\ref{lagid1}-\ref{lagid3}). The possibility to perform these sums explicitly ultimately relies on the following relation satisfied by the generating function (\ref{genLaguerre}):
\beq
G_\alpha(t,x)G_\beta(t,y)=G_{\alpha+\beta+1}(t,x+y).
\eeq
Expanding in powers of $t$, one obtains the summation identity
\beq\label{laguerreid}
\sum_{k=0}^n L^\alpha_k(x) L^{\beta}_{n-k}(y)=L^{\alpha+\beta+1}_ n(x+y).
\eeq
Hence, it follows from (\ref{defS}) that
\beq
\sum_{k=0}^{n+j} S_{njk.n+j-k} = \int\limits_0^\infty d\rho \, e^{-2\rho} \rho^{2\mm} L^\mm_{n}(\rho) L^\mm_j(\rho) L^{2\mm+1}_{n+j}(2\rho).
\eeq
We shall need below the values of Laguerre polynomials and their derivatives at the origin. From (\ref{genLaguerre}),
\begin{align}
&L^\alpha_n(0)=\frac{1}{n!}\del_t^n \frac1{(1-t)^{\alpha+1}}\Bigg|_{t=0}=\frac{\Gamma(n+\alpha+1)}{n!\,\Gamma(\alpha+1)},\label{lag01}\\
&\del_x L^\alpha_n(0)=-\frac{1}{n!}\del_t^n\frac{t}{(1-t)^{\alpha+2}}\Bigg|_{t=0}=-\frac{\Gamma(n+\alpha+1)}{(n-1)!\,\Gamma(\alpha+2)},\label{lag02}\\
&\del_x^2 L^\alpha_n(0)=\frac{1}{n!}\del_t^n\frac{t^2}{(1-t)^{\alpha+3}}\Bigg|_{t=0}=\frac{\Gamma(n+\alpha+1)}{(n-2)!\,\Gamma(\alpha+3)}\label{lag03}.
\end{align}
From (\ref{lag01}), we infer that
\beq
L^\mm_{n}(\rho) L^\mm_j(\rho) =\frac{\Gamma(n+\mm+1)\Gamma(j+\mm+1)}{n!j!(\Gamma(\mm+1))^2} +\rho P_{n+j-1}(\rho),
\eeq
where $P_{n+j-1}$ is a polynomial of degree $n+j-1$. Since $L^{2\mm+1}_n$ are orthogonal with respect to the measure $\rho^{2\mm+1} e^{-\rho}$, one concludes that
\beq\label{lagsumint}
\sum_{k=0}^{n+j} S_{njk.n+j-k} = \frac{\Gamma(n+\mm+1)\Gamma(j+\mm+1)}{n!j!(\Gamma(\mm+1))^2}\int\limits_0^\infty d\rho \, e^{-2\rho} \rho^{2\mm} L^{2\mm+1}_{n+j}(2\rho).
\eeq
The following integration identity can be derived from (\ref{genLaguerre}):
\beq\label{intLaguerre}
\int\limits_0^\infty d\rho \, e^{-\rho}  \rho^{\beta} L^\alpha_n(\rho)=\frac{1}{n!}\del_t^n\int\limits_0^\infty d\rho \, \rho^{\beta}\frac{e^{-\frac{\rho}{1-t}}}{(1-t)^{\alpha+1}}\Bigg|_{t=0}\hspace{-3mm}=\frac{\Gamma(\beta+1)}{n!}\del_t^n\frac1{(1-t)^{\alpha-\beta}}\Bigg|_{t=0}\hspace{-3mm}=\frac{\Gamma(\beta+1)\,\Gamma(\alpha-\beta+n)}{n!\,\Gamma(\alpha-\beta)}.
\eeq
Once the integral in (\ref{lagsumint}) has been thus evaluated, one obtains (\ref{lagid1}).

The next identity we have to prove, which is (\ref{lagid2}), in fact follows from (\ref{lagid1}) if one recalls the symmetry of $S$ under permutation of the third and fourth indices \cite{AO}. We shall nonetheless derive it directly in a way that parallels the above derivation of (\ref{lagid1}).

We first notice that
\beq\label{deltG}
t\del_t G_\alpha(t,x)=(\alpha+1)tG_{\alpha+1}-xtG_{\alpha+2}.
\eeq
Hence,
\beq
[t\del_t G_\alpha(t,x)] G_\beta(t,y)=(\alpha+1)tG_{\alpha+\beta+2}(t,x+y)-xtG_{\alpha+\beta+3}(t,x+y),
\eeq
and
\beq\label{laguerrekid}
\sum_{k=0}^n kL^\alpha_k(x)L^\beta_{n-k}(y)=(\alpha+1)L^{\alpha+\beta+2}_{n-1}(x+y)-xL^{\alpha+\beta+3}_{n-1}(x+y).
\eeq
Therefore,
\beq
\sum_{k=0}^{n+j} k S_{njk.n+j-k}=  \int\limits_0^\infty d\rho \, e^{-2\rho} \rho^{2\mm} L^\mm_{n}(\rho) L^\mm_j(\rho) \left[(\mm+1)L^{2\mm+2}_{n+j-1}(2\rho)-\rho L^{2\mm+3}_{n+j-1}(2\rho)\right].
\eeq
From (\ref{lag01}-\ref{lag02}),
\beq
L^\mm_{n}(\rho) L^\mm_j(\rho)=\frac{\Gamma(n+\mm+1)\Gamma(j+\mm+1)}{n!j!(\Gamma(\mm+1))^2}\left(1-\frac{n+j}{\mm+1}\rho\right)+\rho^2P_{n+j-2}(\rho),
\eeq
where $P_{n+j-2}$ is a polynomial of degree $n+j-2$. Since  $L^{\alpha}_n$ are orthogonal with respect to the measure $\rho^{\alpha} e^{-\rho}$, one obtains
\begin{align}
\sum_{k=0}^{n+j} k S_{njk.n+j-k}=& \frac{\Gamma(n+\mm+1)\Gamma(j+\mm+1)}{n!j!(\Gamma(\mm+1))^2} \\
&\nonumber\times\int\limits_0^\infty d\rho \, e^{-2\rho} \rho^{2\mm} \left(1-\frac{n+j}{\mm+1}\rho\right) \left[(\mm+1)L^{2\mm+2}_{n+j-1}(2\rho)-\rho L^{2\mm+3}_{n+j-1}(2\rho)\right].
\end{align}
With (\ref{intLaguerre}), this is evaluated to yield (\ref{lagid2}).

To prove the last summation identity (\ref{lagid3}), we observe that
\begin{align}
[t\del_t G_\alpha(t,x)] [t\del_t G_\beta(t,y)]=&(\alpha+1)(\beta+1)t^2G_{\alpha+\beta+3}(x+y)\\
&-[(\beta+1)x+(\alpha+1)y]t^2G_{\alpha+\beta+4}(x+y)+xyt^2G_{\alpha+\beta+5}(x+y).\nonumber
\end{align}
Expanding in powers of $t$ gives
\interdisplaylinepenalty=10000
\begin{align}\label{laguerrek2id}
\sum_{k=0}^n k(n-k)L^\alpha_k(x)L^\beta_{n-k}(y)=&(\alpha+1)(\beta+1)L^{\alpha+\beta+3}_{n-2}(x+y)\\
&-[(\beta+1)x+(\alpha+1)y]L^{\alpha+\beta+4}_{n-2}(x+y)+xyL^{\alpha+\beta+5}_{n-2}(x+y).\nonumber
\end{align}
\interdisplaylinepenalty=0
Hence,
\begin{align}\label{lagint2}
\sum_{k=0}^{n+j}k(n+j-k)S_{njk.n+j-k}=&    \int\limits_0^\infty d\rho \, e^{-2\rho} \rho^{2\mm} L^\mm_{n}(\rho) L^\mm_j(\rho)\\
&\times\left[(\mm+1)^2L^{2\mm+3}_{n+j-2}(2\rho)-2\rho(\mm+1) L^{2\mm+4}_{n+j-2}(2\rho)+\rho^2 L^{2\mm+5}_{n+j-2}(2\rho)\right].\nonumber
\end{align}
From (\ref{lag01}-\ref{lag03}),
\begin{align}
 L^\mm_{n}(\rho) L^\mm_j(\rho)=\frac{\Gamma(n+\mm+1)\Gamma(j+\mm+1)}{n!j!(\Gamma(\mm+1))^2}&\left(1-\frac{n+j}{\mm+1}\rho+\left(\frac{nj}{(\mm+1)^2}+\frac{n(n-1)+j(j-1)}{2(\mm+1)(\mm+2)}\right)\rho^2\right)\nonumber\\
&+\rho^3P_{n+j-3}(\rho)\rule{0mm}{7mm},
\end{align}
where $P_{n+j-3}$ is a polynomial of degree $n+j-3$.
Substituting this in (\ref{lagint2}), dropping all terms involving $P_{n+j-3}$ on account of orthogonality, and evaluating by (\ref{intLaguerre}) gives (\ref{lagid3}).\vspace{5mm}

\section{Appendix B: Identities for the interaction coefficients of Landau level truncations}

We start by establishing (\ref{G_L}) as a consequence of (\ref{genLaguerre}):
\begin{align}
&G_L(s,\rho)=\sum_{n=0}^\infty \frac{s^n}{n!} L_L^{n-L}(\rho)
=\frac1{L!}\del_t^L\sum_{n=0}^\infty \frac{s^n}{n!}
\frac{e^{-\frac{t\rho}{1-t}}}{(1-t)^{n-L+1}}\Bigg|_{t=0}\\
&\hspace{3cm}=\frac{e^s}{L!}\del_t^L\Big((1-t)^{L-1}e^{-\frac{t(\rho-s)}{1-t}}\Big)\Big|_{t=0}=\frac{e^s(s-\rho)^L}{L!}.\nonumber
\end{align}
The last equality can be obtained by first noticing that $\del_t^L\big((1-t)^{L-1}e^{-\frac{tx}{1-t}}\big)$ at $t=0$ is a polynomial of degree $L$ in $x$. Computing its derivatives at $x=0$ straightforwardly shows that the first $L-1$ derivatives vanish, while the $L$th derivative is $L!(-1)^L$. Hence the polynomial is simply $(-x)^L$, which gives the desired result.

From the above expression,
\beq
\sum_{k=0}^N
\frac{L_L^{k-L}(\rho)L_L^{N-k-L}(\rho)}{k!(N-k)!}=\frac1{N!}\del_s^N [G_L(s,\rho)]^2\Big|_{s=0}=\frac1{N!(L!)^2}\del_s^N\left(e^{2s}(s-\rho)^{2L}\right)\Big|_{s=0}.
\eeq
Hence,
\beq
\sum_{k=0}^{n+m} \sqrt{\frac{n!m!}{k!(n+m-k)!}}\tildeC_{nmk,n+m-k}= \int\limits_{0}^{\infty}d\rho\, \rho^{n+m-2L} L_{L}^{n-L}(\rho)L_{L}^{m-L}(\rho)\frac1{(n+m)!}\del_s^{n+m}\left(e^{2(s-\rho)}(s-\rho)^{2L}\right)\Big|_{s=0}.\nonumber
\eeq
Since the $s$-derivatives act on an expression that only depends on $s-\rho$, each $\del_s$ can be identically traded for $-\del_\rho$, after which $s$ is straightforwardly set to 0 (independently of the remaining $\rho$-differentiations). This yields
\beq
\sum_{k=0}^{n+m} \sqrt{\frac{n!m!}{k!(n+m-k)!}}\tildeC_{nmk,n+m-k}= \int\limits_{0}^{\infty}d\rho\, \rho^{n+m-2L} L_{L}^{n-L}(\rho)L_{L}^{m-L}(\rho)\frac{(-1)^{n+m}}{(n+m)!}\del_\rho^{n+m}\left(e^{-2\rho}\rho^{2L}\right).
\eeq
Integrating by parts $(n+m)$ times gives
\beq
\sum_{k=0}^{n+m} \sqrt{\frac{n!m!}{k!(n+m-k)!}}\tildeC_{nmk,n+m-k}= \int\limits_{0}^{\infty}d\rho\, e^{-2\rho}\rho^{2L}\frac{1}{(n+m)!}\del_\rho^{n+m}\left( \rho^{n+m-2L} L_{L}^{n-L}(\rho)L_{L}^{m-L}(\rho)\right).
\eeq
The $(n+m)$ derivatives now act on a polynomial of degree $(n+m)$ in $\rho$, hence the only nonvanishing contribution can come from the highest power of $\rho$, which in turn comes from the highest powers of $\rho$ in $L_{L}^{n-L}(\rho)$ and $L_{L}^{m-L}(\rho)$, which are both $(-1)^L\rho^L/L!$, independently of $n$ and $m$, as one can read off (\ref{Lag_explicit}). Hence,
\beq
\sum_{k=0}^{n+m} \sqrt{\frac{n!m!}{k!(n+m-k)!}}\tildeC_{nmk,n+m-k}= \int\limits_{0}^{\infty}d\rho\, e^{-2\rho}\rho^{2L}\frac{1}{(n+m)!(L!)^2}\del_\rho^{n+m}\left( \rho^{n+m}\right)=\frac{1}{(L!)^2}\int\limits_{0}^{\infty}d\rho\,e^{-2\rho}\rho^{2L}.\nonumber
\eeq
Evaluating the last integral gives (\ref{LLid1}).

The next identity (\ref{LLid2}) in fact follows from (\ref{LLid1}) due to the symmetries of $S$ \cite{AO}, but we shall prove it explicitly for completeness. One has
\beq
\sum_{k=0}^N k
\frac{L_L^{k-L}(\rho)L_L^{N-k-L}(\rho)}{k!(N-k)!}=\frac1{N!}\del_s^N [G_L(s,\rho)s\del_s G_L(s,\rho)]\Bigg|_{s=0}\hspace{-3mm}=\frac1{N!(L!)^2}\del_s^N\left(e^{2s}(s-\rho)^{2L-1}s(L+s-\rho)\right)\Bigg|_{s=0}\nonumber
\eeq
and
\begin{align}
&\sum_{k=0}^{n+m} k \sqrt{\frac{n!m!}{k!(n+m-k)!}}\tildeC_{nmk,n+m-k}\\
&\hspace{2cm}= \int\limits_{0}^{\infty}d\rho\, \frac{\rho^{n+m-2L}} {(n+m)!}L_{L}^{n-L}(\rho)L_{L}^{m-L}(\rho)\del_s^{n+m}\left([(s-\rho)+\rho] e^{2(s-\rho)}(s-\rho)^{2L-1}(L+s-\rho)\right)\Big|_{s=0}.\nonumber
\end{align}
The strategy is to remove all $\rho$-factors outside the $\del_s$-derivatives, so that the latter act on an expression depending exclusively of $s-\rho$ and can be traded for $-\del_\rho$, as in our proof of the first identity. This yields
\begin{align}
&\sum_{k=0}^{n+m} k \sqrt{\frac{n!m!}{k!(n+m-k)!}}\tildeC_{nmk,n+m-k}\\
&\hspace{3cm}= \int\limits_{0}^{\infty}d\rho \,\rho^{n+m-2L} L_{L}^{n-L}(\rho)L_{L}^{m-L}(\rho)\frac{(-1)^{n+m}}{(n+m)!}\del_\rho^{n+m}\left(e^{-2\rho}\rho^{2L}(L-\rho)\right)\nonumber\\
&\hspace{3cm}-\int\limits_{0}^{\infty}d\rho\, \rho^{n+m-2L+1} L_{L}^{n-L}(\rho)L_{L}^{m-L}(\rho)\frac{(-1)^{n+m}}{(n+m)!}\del_\rho^{n+m}\left(e^{-2\rho}\rho^{2L-1}(L-\rho)\right).\nonumber
\end{align}
We need terms of degrees $2L$ and $2L-1$ in the product $L_{L}^{n-L}(\rho)L_{L}^{m-L}(\rho)$, which are
\beq
L_{L}^{n-L}(\rho)L_{L}^{m-L}(\rho)=\frac{\rho^{2L-1}}{(L!)^2}\left(\rho-(n+m)L \right)+\cdots,
\eeq
as follows from (\ref{Lag_explicit}). Then, integrating by parts $(n+m)$ times and evaluating the remaining elementary integrals yields (\ref{LLid2}).

For the last identity (\ref{LLid3}), we write
\beq
\sum_{k=0}^N k(N-k)
\frac{L_L^{k-L}(\rho)L_L^{N-k-L}(\rho)}{k!(N-k)!}=\frac1{N!}\del_s^N [s\del_s G_L(s,\rho)]^2\Bigg|_{s=0}\hspace{-3mm}=\frac1{N!(L!)^2}\del_s^N\left(s^2 e^{2s}(s-\rho)^{2L-2}(L+s-\rho)^2\right)\Bigg|_{s=0}.\nonumber
\eeq
Hence,
\begin{align}
&\sum_{k=0}^{n+m} k(n+m-k) \sqrt{\frac{n!m!}{k!(n+m-k)!}}\tildeC_{nmk,n+m-k}\\
&\hspace{1mm}= \int\limits_{0}^{\infty}d\rho\, \frac{\rho^{n+m-2L}} {(n+m)!}L_{L}^{n-L}(\rho)L_{L}^{m-L}(\rho)\del_s^{n+m}\left([(s-\rho)^2+2\rho(s-\rho)+\rho^2] e^{2(s-\rho)}(s-\rho)^{2L-2}(L+s-\rho)^2\right)\Big|_{s=0}.\nonumber
\end{align}
By processing identical to the above examples,
\begin{align}
&\sum_{k=0}^{n+m} k(n+m-k) \sqrt{\frac{n!m!}{k!(n+m-k)!}}\tildeC_{nmk,n+m-k}\\
&\hspace{3cm}= \int\limits_{0}^{\infty}d\rho \,\rho^{n+m-2L} L_{L}^{n-L}(\rho)L_{L}^{m-L}(\rho)\frac{(-1)^{n+m}}{(n+m)!}\del_\rho^{n+m}\left(e^{-2\rho}\rho^{2L}(L-\rho)^2\right)\nonumber\\
&\hspace{3cm}-2\int\limits_{0}^{\infty}d\rho \,\rho^{n+m-2L+1} L_{L}^{n-L}(\rho)L_{L}^{m-L}(\rho)\frac{(-1)^{n+m}}{(n+m)!}\del_\rho^{n+m}\left(e^{-2\rho}\rho^{2L-1}(L-\rho)^2\right)\nonumber\\
&\hspace{3cm}+\int\limits_{0}^{\infty}d\rho\, \rho^{n+m-2L+2} L_{L}^{n-L}(\rho)L_{L}^{m-L}(\rho)\frac{(-1)^{n+m}}{(n+m)!}\del_\rho^{n+m}\left(e^{-2\rho}\rho^{2L-2}(L-\rho)^2\right).\nonumber
\end{align}
One again integrates by parts $(n+m)$ times, but now we need terms of degree $2L$, $2L-1$ and $2L-2$ in the product $L_{L}^{n-L}(\rho)L_{L}^{m-L}(\rho)$, which are extracted from (\ref{Lag_explicit}) as
\beq
L_{L}^{n-L}(\rho)L_{L}^{m-L}(\rho)=\frac{\rho^{2L-2}}{(L!)^2}\left(\rho^2-(n+m)L\rho+\frac{L(L-1)}{2}[n^2+m^2-(n+m)]+L^2nm\right)+\cdots
\eeq
Putting everything together yields (\ref{LLid3}).

\twocolumngrid


\begin{thebibliography}{99}

\bibitem{BDZ}I.~Bloch, J.~Dalibard and W.~Zwerger, \emph{Many-body physics with ultracold gases}, Rev. Mod. Phys. {\bf 80} (2008) 885 \arXiv{0704.3011} [cond-mat.other].

\bibitem{cooper}N.~R.~Cooper, \emph{Rapidly rotating atomic gases}, Adv. Phys. 57 (2008) 539 \arXiv{0810.4398} [cond-mat.mes-hall].

\bibitem{fetter}A.~L.~Fetter, \emph{Rotating trapped Bose-Einstein condensates}, Rev. Mod. Phys. {\bf 81} (2009) 647  \arXiv{0801.2952} [cond-mat.stat-mech].

\bibitem{GHT} P. Germain, Z. Hani and L. Thomann,
    \emph{On the continuous resonant equation for NLS: I. Deterministic analysis,} J. Math. Pur. App. {\bf 105} (2016) 131
    \arXiv{1501.03760} [math.AP].

\bibitem{BBCE}
  A.~Biasi, P.~Bizo\'n, B.~Craps and O.~Evnin,
  {\em Exact lowest-Landau-level solutions for vortex precession in Bose-Einstein condensates,}
  Phys.\ Rev.\ A {\bf 96} (2017) 053615
  \arXiv{1705.00867} [cond-mat.quant-gas].

\bibitem{GT} P.~Germain and L.~Thomann,  \emph{On the high frequency limit of the LLL equation}, Quart. Appl. Math. {\bf 74} (2016) 633 \arXiv{1509.09080} [math.AP].

\bibitem{GGT}
  P.~G\'erard, P.~Germain and L.~Thomann,
  {\em On the cubic lowest Landau level equation,}
  \arXiv{1709.04276} [math.AP].

\bibitem{BMP}A.~F.~Biasi, J.~Mas and A.~Paredes,
  \emph{Delayed collapses of BECs in relation to AdS gravity,}
  Phys.\ Rev.\ E {\bf 95} (2017) 032216 \arXiv{1610.04866} [nlin.PS].

\bibitem{FPU}
  V.~Balasubramanian, A.~Buchel, S.~R.~Green, L.~Lehner and S.~L.~Liebling,
  {\em Holographic thermalization, stability of anti-de Sitter space, and the Fermi-Pasta-Ulam paradox,}
  Phys.\ Rev.\ Lett.\  {\bf 113} (2014) 071601
  \arXiv{1403.6471} [hep-th].

\bibitem{CEV1} B.~Craps, O.~Evnin and J.~Vanhoof,
 \emph{Renormalization group, secular term resummation and AdS (in)stability,}
 JHEP {\bf 1410} (2014) 48
 \arXiv{1407.6273} [gr-qc].

\bibitem{CEV2}
B.~Craps, O.~Evnin and J.~Vanhoof,
\emph{Renormalization, averaging, conservation laws and AdS (in)stability,}
JHEP {\bf 1501} (2015) 108
\arXiv{1412.3249} [gr-qc].

\bibitem{CF}  P.~Bizo\'n, B.~Craps, O.~Evnin, D.~Hunik, V.~Luyten and M.~Maliborski,
 \emph{Conformal flow on $S^3$ and weak field integrability in AdS$_4$,} Comm.\ Math.\ Phys.\  {\bf 353} (2017) 1179 \arXiv{1608.07227} [math.AP].

\bibitem{BEL}
  B.~Craps, O.~Evnin and V.~Luyten,
  {\em Maximally rotating waves in AdS and on spheres,} JHEP {\bf 1709} (2017) 059
   \arXiv{1707.08501} [hep-th].

\bibitem{BR} P.~Bizo\'n and A.~Rostworowski,
 {\em On weakly turbulent instability of anti-de Sitter space,}
 Phys.\ Rev.\ Lett.\ {\bf 107} (2011) 031102
 \arXiv{1104.3702} [gr-qc].

\bibitem{BMR} P.~Bizo\'n, M.~Maliborski, A.~Rostworowski, \emph{Resonant dynamics and the instability of anti-de Sitter spacetime,} Phys.\ Rev.\ Lett. {\bf 115} (2015) 081103
    \arXiv{1506.03519} [gr-qc].

\bibitem{rev2} B.~Craps and O.~Evnin,
 {\em AdS (in)stability: an analytic approach,}
Fortsch.\ Phys.\ {\bf 64} (2016) 336
 \arXiv{1510.07836} [gr-qc].

\bibitem{AO} A.~Biasi, P.~Bizo\'n and O.~Evnin, {\em Solvable cubic resonant systems},  \arXiv{1805.03634} [nlin.SI].

\bibitem{GG}P.~G\'erard and S.~Grellier,
{\em The cubic Szeg\H o equation,} Ann. Scient. \'Ec. Norm. Sup. {\bf 43} (2010) 761
\arXiv{0906.4540} [math.CV].

\bibitem{Niederer}
U.~Niederer, {\em The maximal kinematical invariance group of the harmonic oscillator,} Helv. Phys. Acta 46 (1973) 191.

\bibitem{OFN}
  K.~Ohashi, T.~Fujimori and M.~Nitta,
  {\em Conformal symmetry of trapped Bose-Einstein condensates and massive Nambu-Goldstone modes,}
  Phys.\ Rev.\ A {\bf 96} (2017) 051601
  \arXiv{1705.09118} [cond-mat.quant-gas].

\bibitem{translations}J.~J.~Garc\'\i a-Ripoll, V.~M.~P\'erez-Garc\'\i a and V.~Vekslerchik, {\it Construction of exact solutions by spatial translations in inhomogeneous non\-linear Schr\"odinger equations}, Phys.\ Rev.\ E {\bf 64} (2001) 056602.

\bibitem{dahl}J.~P.~Dahl and W.~P.~Schleich, \emph{State operator, constants of the motion, and Wigner functions: The two-dimensional
isotropic harmonic oscillator},  Phys.\ Rev.\ A {\bf 79}  (2009) 024101.

\bibitem{murdock}  J.~A.~Murdock, \emph{Perturbations: Theory and Methods}, SIAM (1987).

\bibitem{KM} S.~Kuksin and A.~Maiocchi, {\em The effective equation method}, in {\em New Approaches to Nonlinear Waves}, Springer (2016) \arXiv{1501.04175} [math-ph].

\bibitem{dong} S.-H.~Dong, {\em Factorization Method in Quantum Mechanics}, Springer (2007).

\bibitem{BHP1} P.~Bizo\'n, D.~Hunik-Kostyra and D.~Pelinovsky, {\em Ground state of the conformal flow on $S^3$}, \arXiv{1706.07726} [math.AP], to appear in Comm. Pure Appl. Math.

\bibitem{BHP2} D.~Pelinovsky, D.~Hunik-Kostyra and P.~Bizo\'n, {\em Stationary states of the cubic conformal flow on $S^3$}, \arXiv{1807.00426} [math-ph].

\bibitem{Kevrekidis} P.~G.~Kevrekidis, D.~J.~Frantzeskakis and R.~Carretero-Gonz\'alez (eds), \emph{Emergent nonlinear phenomena in Bose-Einstein condensates}, Springer (2008).

\bibitem{PK} D.~E.~Pelinovsky and  P.~G.~Kevrekidis, {\em Periodic oscillations of dark solitons in parabolic potentials,} AMS Cont. Math. {\bf 473} (2008) 159 \arXiv{0705.1016} [cond-mat.other].

\bibitem{Wang} W.~Wang, P.~G.~Kevrekidis, R.~Carretero-Gonz\'alez and D.~J.~Frantzeskakis, \emph{Dark spherical shell solitons in three-dimensional Bose-Einstein condensates: Existence, stability and dynamics},  Phys.\ Rev.\ A {\bf 93} (2016) 023630 \arXiv{1601.02176} [cond-mat.quant-gas].

\bibitem{ring1}D.~J.~Frantzeskakis, \emph{Dark solitons in atomic Bose-Einstein condensates: from theory to experiments,} J.\ Phys.\ A {\bf 43} (2010) 213001 \arXiv{0709.2132} [quant-ph].

\bibitem{ring2}T.~Kapitula, P.~G.~Kevrekidis and R.~Carretero-Gonz\'alez, \emph{Rotating matter waves in Bose-Einstein condensates,} Physica\ D {\bf 233} (2007) 112.

\bibitem{ring3}E.~G.~Charalampidis, P.~G.~Kevrekidis and P.~E.~Farrell, \emph{Computing stationary solutions of the two-dimensional Gross-Pitaevskii equation with deflated continuation,} 
Comm. Nonlin. Sci. Num. Sim. 54 (2018) 482 \arXiv{1612.08145} [nlin.PS].

\bibitem{arr1}A.~Klein, D.~Jaksch, Y.~Zhang and W.~Bao, \emph{Dynamics of vortices in weakly interacting Bose-Einstein condensates,} Phys.\ Rev.\ A {\bf 76} (2007) 043602 \arXiv{0709.2132} [quant-ph].

\bibitem{arr2}W.~Li, M.~Haque and S.~Komineas, \emph{A vortex dipole in a trapped two-dimensional Bose-Einstein condensate,} Phys.\ Rev.\ A {\bf 77} (2008) 053610 \arXiv{0712.4217} [cond-mat.other].

\bibitem{arr3}P.~J.~Torres, R. Carretero-Gonz\'alez, S.~Middelkamp, P.~Schmelcher,
D.~J.~Frantzeskakis and P.~G.~Kevrekidis, \emph{Vortex interaction dynamics in trapped Bose-Einstein condensates,} Comm. Pur. App. Analysis {\bf 10} (2011) 1589.

\bibitem{arrexp}J.~A.~Seman, E.~A.~L.~Henn, M.~Haque, R.~F.~Shiozaki, E.~R.~F.~Ramos, M.~Caracanhas, C.~Castelo Branco, P.~E.~S.~Tavares, F.~J.~Poveda-Cuevas, G.~Roati, K.~M.~F.~Magalh\~aes and V. S. Bagnato, \emph{Three-vortex configurations in trapped Bose-Einstein condensates, } Phys. Rev. A {\bf 82} (2010) 033616  	\arXiv{0907.1584} [cond-mat.quant-gas].

\end{thebibliography}
\end{document}